\documentclass[aip,jcp,numerical,reprint,floatfix]{revtex4-1}

\usepackage{graphicx}
\usepackage{dcolumn}
\usepackage{bm}
\usepackage{amsmath}
\usepackage{multirow}
\usepackage{xcolor}

\newcolumntype{f}[1]{D{.}{.}{#1}}

\begin{document}

\title{Shake-Rattle-and-Roll: A Model of Dynamic Structural Disorder in
Supported Nanoscale Catalysts}

\author{J. J. Rehr and F. D. Vila}
\affiliation{Department of Physics, University of Washington, Seattle, WA 98195}

\begin{abstract}

We investigate the effects of ``dynamic structural disorder" (DSD) on the
behavior of supported nano-scale catalysts. DSD refers to the intrinsic
fluctuating, inhomogeneous structure of such nano-scale systems.  In contrast to
bulk materials, nano-scale systems exhibit substantial fluctuations in energy,
charge, and other extensive quantities  as well as large surface effects. The
DSD is driven by the stochastic librational motion of the center of mass and
fluxional bonding at the nanoparticle surface due to thermal coupling with the
substrate. Our approach for calculating DSD  is based on a combination of
statistical mechanics, transient coupled-oscillator models, and real-time DFT/MD
simulations. This approach treats thermal and dynamic effects over multiple
time-scales, including bond-stretching and -bending vibrations, DSD, and
transient tethering to the substrate at longer ps time-scales.  Model
calculations of molecule-cluster interactions and molecular dissociation
reaction paths are presented in which the reactant molecules are adsorbed on the
surface of dynamically sampled clusters.  This model suggests that DSD affects
both the prefactors and distribution of energy barriers in reaction rates, and
thus can strongly affect catalytic activity at the nano-scale.

\end{abstract}

\date{\today}
\maketitle

\section{Introduction}

 The structure and behavior of materials at the nanoscale are both of
fundamental and technological importance
and are
particularly relevant to the understanding  of 
supported nanoscale catalysts. This challenging problem is
complicated by the fact
that nano-scale properties differ from those of 
condensed systems, especially at high temperature.
 Their unusual thermodynamic properties have been widely
recognized at the structural,\cite{Billinge27042007}
catalytic,\cite{Johanek11062004} and spectroscopic\cite{link2003optical}
level. They differ from the bulk since the thermodynamic limit 
does not apply for small $N$, where $N$ refers to the number of atoms
in the nanocluster.  For example, cluster surface effects of $O(N^{2/3})$
are crucial and thus nano-scale systems are
necessarily inhomogeneous.  Moreover, statistical physics arguments imply
that nano-scale systems exhibit substantial $O(N^{1/2})$ fluctuations
in extensive physical quantities like total energy and charge,
and hence $O(N^{-1/2})$ fluctuations in local temperature. In addition
supported nano-scale catalysts exhibit an intrinsic fluctuating,
inhomogeneous structure referred to as dynamic structural disorder
(DSD).\cite{vila2013} Thus, in contrast to bulk materials, they have
no well-defined ``equilibrium structure," particularly at high
temperatures. The presence of these intrinsic fluctuations suggest
that it would be useful to examine their behavior from a dynamical perspective.
This approach is in contrast to conventional surface science approaches
typically based on time-independent
structural properties.
Indeed, there has been a growing interest in such
real-time approaches 
in recent years.\cite{kang2006,PhysRevB.78.121404,vila2013,C3SC50614B}
Dynamic effects have
also been postulated to play a role in other treatments of catalysis,
ranging from heterogeneous catalysis\cite{Topsoe2003155,boero1998}
at surfaces and in the catalytic activity in biostructures.
\cite{schramm2007enzymatic,min2005}
The difference in the treatment here is the emphasis on the effects of
nano-scale dynamics on catalytic properties which are investigated using
nudged-elastic-band (NEB) transition state theory (TST) calculations.

Our approach for
modelling the effects of DSD
is based on a
concept dubbed picturesquely as ``Shake-Rattle-and-Roll" (SRR).
This approach was inspired by observations
of unusual behavior in nanoscale supported Pt and Pt-alloy catalysts:
In particular, x-ray absorption spectra (XAS) studies showed 
that small Pt nano-clusters exhibit negative
thermal expansion, anomalously large disorder, and
temperature dependent shifts in XAS threshold energies.\cite{kang2006}
 These anomalous
properties were subsequently explained by finite-temperature DFT/MD
calculations, which showed that they are dynamical in origin, involving
multiple time-scales.\cite{PhysRevB.78.121404}
They include 1) fast bond
vibrations (i.e., {\it shaking})
2) soft anharmonic or fluxional modes, particularly
involving the nano-particle surface atoms; and
3) librational motion of the center of mass (CM),
which is due to the thermal contact of the nano-clusters
tethered to the support bonding sites.  As discussed below, these CM dynamics
are analogous to hindered 2-dimensional Brownian motion.
When combined with the stochastic thermal coupling to the support,
these low-frequency modes induce a {\it rattle-}like motion
in the system, typically at sub-THz frequencies.
At longer time scales (tens of ps for supported Pt clusters on
$\gamma$-alumina at high temperatures), bonding to the support is
transient. When those bonds break, the nano-clusters
can then {\it roll} or slide to new positions on the support.
The combined motion of many such clusters eventually leads to sintering,
which typically reduces catalytic activity.  
This rattle-like motion has recently been interpreted as
dynamic structural disorder, in contrast to
static
bond distortions
or substitutional disorder typical in condensed systems.
Instead, the DSD  is ``non-equilibrium" in character, in the sense that
the system exhibits fluctuations from from mechanical and thermal
equilibrium,
even though, of course, the full system (cluster plus support) is
in overall ``thermal-equilibrium," with a well defined
global temperature $T$.
Thus the geometric structure of nano-clusters is amorphous and fluctuating,
rather like that in a partly-melted solid. Indeed, DSD affects the
nano-cluster behavior through stochastic fluctuations in structure,
cluster charge and center of mass position.

These remarkable observations have motivated us to
reinvestigate the structure and catalytic behavior of supported nanoscale
catalysts in terms of their dynamic structure.  To this end, we
introduce the SRR concept based on a combination of statistical mechanics,
transient coupled-oscillator models, and real-time DFT/MD simulations.
As discussed below,
SRR
can explain the anomalous properties
observed for supported nano-clusters, including both DSD and charge
fluctuations.  As an illustration we have applied
this approach to
calculations of
both the interaction between the supported nanocatalysts and
prototypical molecules, and their dissociation reaction
paths.
These show that the DSD provides a mechanism
for increasing reaction rates.  Although still preliminary, the SRR model
offers tantalizing insights into dynamic mechanisms that contribute to
catalytic processes.

\section{Thermal Fluctuations at the Nanoscale}

 As noted in the introduction, thermal properties of nano-structures with
relatively small $N$ differ substantially from those in macroscopic
condensed matter due to finite-size effects enhanced by their poor thermal
coupling to the support. In contrast to bulk systems, their mean energy
$\bar E$ is not sharply defined, but exhibits substantial
energy fluctuations of order $kT \sqrt{N}$,
\cite{landau1980statistical}  leading to fluctuations in the local
temperature 
(or energy per particle) of order $T/\sqrt{N}$. This effect is not
immediately obvious since the global temperature $T$ is a constant
throughout the system.  To understand these results, we recount the
classical arguments of statistical thermodynamics.
We consider a nano-cluster with $N$ atoms weakly bound to a support
at fixed temperature $T$, which serves as a heat-bath. For simplicity of
argument the cluster volume $V$, charge $Q$, and composition can be
regarded as fixed, but these contraints are not essential and will be
relaxed below.  Due to contact with the support, which is
in continuous thermal and vibrational motion, energy will fluctuate between
the cluster and the support, with a probability distribution
\begin{equation}
 P(E,T) \approx \Omega(E) e^{-\beta E} = e^{-\beta {\cal F}(E,T)}
\end{equation}
where ${\cal{ F}}(E,T)=E-TS(E)$ and
$k\ln \Omega(E)= S(E) $ is the net cluster entropy due to distinguishable
configurations $\Omega(E)$ of the nano-cluster at energy $E$.
Thermal equilibrium corresponds to the maximum probability or, equivalently,
$\min{\cal{F}}(\bar E,T)=F(T)$, i.e., the Helmholtz free energy.
The mean total cluster energy $\bar E$ is then fixed by the relation
$\partial S(E)/\partial E|_{\bar E} = 1/T$, and hence both the nano-cluster
and the support have the same equilibrium temperature $T$.
The cluster entropy $S(E)$ is additive in terms of the independent
$6N$ degrees of freedom in the cluster dynamics.  In addition to the
$(3N-3)$ internal vibrational modes (including both potential and
kinetic degrees of freedom), the dynamics includes four additional
degrees of freedom from the 2-d librational modes
of the CM parallel to the support, and additional modes binding the
cluster to it.  The distribution  $P(E,T)$ is sharply peaked for large $N$
and approximately Gaussian near the peak at $\bar E$.
The  mean square fluctuations $\sigma^2_E$ 
are obtained from the 2nd derivative of the entropy
$\sigma_E^2 = k/[\partial^2 S(E)/\partial E^2]$. The quantity 
$\sigma^2_E$ is clearly $O(N)$ since both the total energy $E$ and 
entropy $S(E)$ of the cluster are extensive.  At high temperatures,
where equipartition is valid,
$\bar E = 3NkT$ and $S(E) \approx 3N k \ln E$.
Thus the energy fluctuations are given by
$\sigma_E \propto kT \sqrt{3N} $ and similarly since $T={\bar E}/3Nk$,
fluctuations in the internal temperature of the cluster are
$\sigma_T = T/\sqrt{3N}$. The effect is rather like that of a fluctuating
thermostat for which any non-linear effects (e.g., reaction rates)
on temperature do not cancel.
Consequently, one may expect substantial
DSD effects on reaction rates.
For example, for $N$=20 at 600 K, $\bar E$ is distributed within a few tenths
of an eV of the mean $\bar E = 3NkT \approx $ 3 eV, and $\sigma_T \approx 75K$. 
In contrast, the thermal fluctuations of the substrate itself are negligible.
This statistical argument suggests why
finite temperature DFT/MD approaches may be more efficient than
full canonical ensemble sampling for calculating the physical properties of
nanoclusters, since they naturally probe the range of
accessible phase space within a few $\sigma_E$ of $\bar E$
over a time-scale comparable to the periods of vibrational and
librational motion.

The above arguments can be generalized for
other conserved physical quantities, for example total volume
$V_{tot}=V+V_S$, net electronic charge $Q_{tot}=Q+Q_S$,
chemical composition, etc., where the subscript $S$ refers to the support.
To this end one must generalize the calculations of accessible states
in terms of $S(E,V,Q,N_i)$ with varying $V$, $Q$, etc.  These additional
physical quantities are stabilized by those from the
support
bath, so that
$F\rightarrow F-pV +\mu Q/e + \Sigma_i \mu_i N_i$ where
$p$ is the pressure, $\mu$ the chemical
potential, $Q/e$ is the number of electrons in the cluster, and 
$N_i$ the number of atoms of species $i$. Since
quantities like $V$ and $Q$ are extensive, one also expects
their fluctuations to be of order
$\sqrt{N}$. The charge (and hence chemical potential) fluctuations,
are important to explain the variation of the Fermi energy with
temperature in the XAS studies.  Due to the strength of Coulomb forces
and the importance of oxidation states in many catalytic
reactions,
these charge fluctuations may play a
key role in controlling
nano-scale properties and chemical reaction rates.  

\section{ Transient coupled oscillator model}

\begin{figure}[ht]
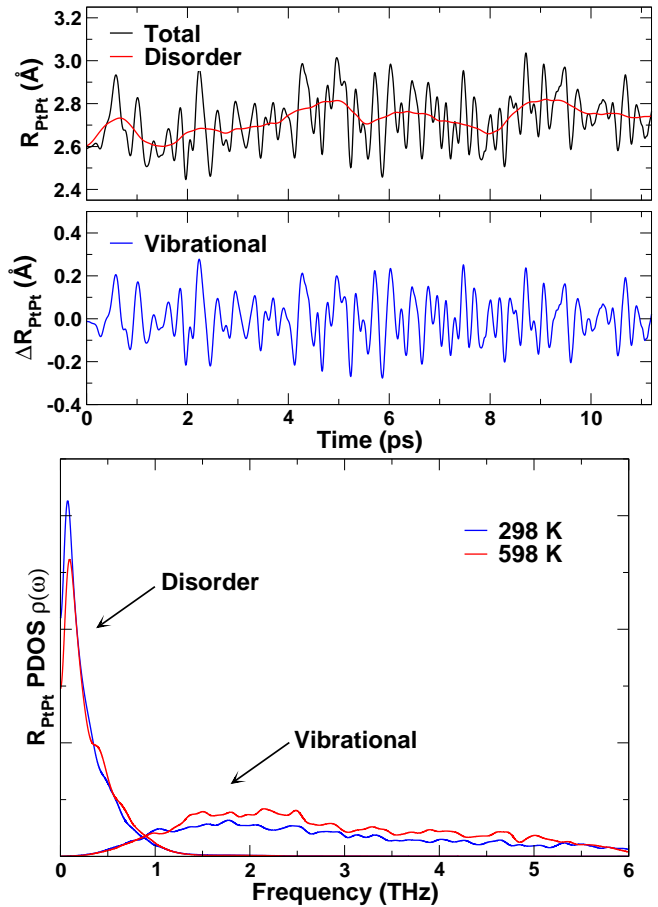

\includegraphics[scale=0.35,clip]{rPtPt_t_Filter.Pt15Sn05.298.eps}
\includegraphics[scale=0.35,clip]{vdos_decomp.Pt15Sn05.298vs598.eps}
\caption{\label{fig:vibdis}
Typical trajectory for a bound Pt-Pt pair in Pt$_{15}$Sn$_{5}$ on
$\gamma$-Al$_{2}$O$_{3}$ at 298 K decomposed into structural disorder and
vibrational components by applying a 0.5 THz low-pass filter (top), and average
projected density of states for the vibrational and dynamic structural disorder components of
bound Pt-Pt pairs in Pt$_{15}$Sn$_{5}$ as a function of temperature (bottom).
}
\end{figure}

The thermodynamic behavior of tethered Pt nano-clusters can be understood
in terms of a ``transient coupled oscillator model." That is, we consider the
internal vibrations with respect to fluctuating instantaneous-equilibrium
structures, which is presumably valid over a ps time-scale long compared
to the  vibrational periods.
The dominant modes of the system coupled to the support thus consist of:
i) vibrational modes; ii) soft and fluxional modes, particularly
close to the cluster surface; and iii) CM librational modes, i.e., soft modes
about which the cluster center of mass fluctuates.  For Pt,
the vibrational modes are of THz scale, while the soft and librational
modes are sub-THz.  The model is transient, since anharmonicity and
bond-breaking can be substantial, and hence the characteristic modes of
oscillation fluctuate.  The thermodynamic quantities of 
interest can be obtained from the free energy e.g., $\langle F(T)\rangle$
averaged over several such models, each of which
can be calculated over sufficiently short (e.g., ps time-scale)
time-intervals within a quasi-harmonic
approximation using the relation,\cite{PhysRevB.76.014301}
\begin{equation}
F(T) = E_0(\bar{R}) + k T \int d\omega\, \rho(\omega)
   \ln [2\sinh (\beta \hbar\omega/2)].
\end{equation}
Here $\rho(\omega)$ is the total density of modes per unit frequency
and $E_0(\bar{R})$ is
the electronic energy at the transient mean atomic positions.
The thermodynamic average $\langle F(T) \rangle$ is given by the same
expression with $E_0(\bar{R})$ replaced by $\langle E_0(\bar{R})\rangle$
and $\rho(\omega)$ by $\langle\rho(\omega)\rangle$.  This approach is efficient,
since despite the fluctuations in structure, the average spectrum
$\langle \rho(\omega)\rangle$ is relatively stable. The term $E_0(\bar R)$ is
important since it contains the transient effects of charge fluctuations
and affects changes in potential energy that dominate the
reaction paths.  The support provides both a heat bath
and charge reservoir at fixed $T$
which give rise to stochastic CM motion and DSD.
 Since we are primarily interested
in catalytic behavior at high-temperatures, we focus our
discussion here to the classical limit 
with mean phonon occupations $n(T) \approx kT/\hbar\omega \gg 1$.
At 600 K this limit is a reasonable approximation for Pt which has a
Debye-temperature of $240$ K and an equivalent Debye frequency of 5.2 THz.
In this limit 
\begin{equation}
 F(T) = E_0(\bar R)
 +   k T \int d\omega\, \rho(\omega) \ln (\beta \hbar\omega).
\end{equation}
This statistical approach can also be directly compared to experiment,
since observed physical properties are usually defined as averages over a
statistical ensemble, and are equivalent to time-averages 
over a long time interval sufficient to cover the accessible phase space.
Low frequency modes are important since $n(T)$ varies inversely with
$\omega$. However, the effect of breathing modes
is expected to be small since for a 3-d system
$\rho(\omega) \approx\omega^2$ at long-wavelengths.
From the identity $\bar E  =\partial \beta F/\partial \beta$, we obtain
a mean energy $\bar E = 3NkT$ at high temperatures, consistent
with the equipartition theorem.  This energy
includes contributions from the 2-d librational motion in the $x-y$ plane,
which can be approximated by
low frequency (sub THz) vibrational modes of the CM (Fig.\ \ref{fig:vibdis}).
Thus the mean
stochastic librational kinetic energy $(1/2)Mv^2_{lib} = kT$. This
result is the analog of Brownian motion, the difference being that 
such motion is hindered and 2-dimensional in character in nanoparticles
tethered to the support.
This stochastic CM motion is crucial to the origin of DSD and
drives non-equlibrium fluctuations in the internal structure.

\section{Dynamic Structural Disorder}

In order to investigate DSD in more detail we consider the effect of
their large fluctuations on the dynamic structure of the cluster.
This structure can be described formally as a point in 6$N$-dimensional phase
space defining the nuclear coordinates and momenta
$\{\vec R_i(t),\vec P_i(t),i=1\dots N\}$.  Due to the fluctuations in the
structure, e.g., the change in shape due to the librational motion and
fluxional surface bonds, the cluster potential energy surface
$V[\{\vec R_i\},t]$ will also fluctuate on a time-scale comparable to
the librational motion, which is much slower than typical bond vibrations.
Thus it is reasonable to assume that a transient harmonic oscillator
approximation is valid for characterizing short-time motion.  During that
time the motion of the cluster is vibrational with respect to the
instantaneous minima on its PES, i.e., fluctuating equilibrium positions,
denoted by $\bar{{R_i}}$.  Next we introduce the mean pair
distribution function (PDF) $g(R)$ defined as the average
distance between all bond pairs  $(i,j)$ in the cluster,
\begin{equation}
g(R) = \frac{1}{N(N-1)}\sum_{i\neq j} \langle
\delta(R - |\vec R_i(t)- \vec R_j(t)|)\rangle.
\end{equation}
This PDF can be measured in experiment. For example the XAFS signal
\begin{equation}
\chi(k) = \int dR\,  g(R)\frac{f_{\rm eff}(k)}{kR^2}
\sin(2kR+\Phi)e^{-2R/\lambda}
\end{equation}
is closely related to the Fourier transform of $g(R)$.  In conventional XAFS
analysis, however, one usually restricts consideration
to the near neighbor bonds, e.g., by Fourier filtering over
the first coordination shell, which is described by the 
near-neighbor distribution function $\tilde g(R)$.
Physical quantities of interest are then
obtained from the cumulant moments of $\tilde g(R)$, e.g. the
mean near-neighbor distance $\bar R = \langle R\rangle =
\int dR\, R\, \tilde g(R)$,
and the mean-square radial disorder (MSRD) by
$\sigma^2 = \langle (R-\bar R)^2\rangle$.

\begin{figure}[ht]
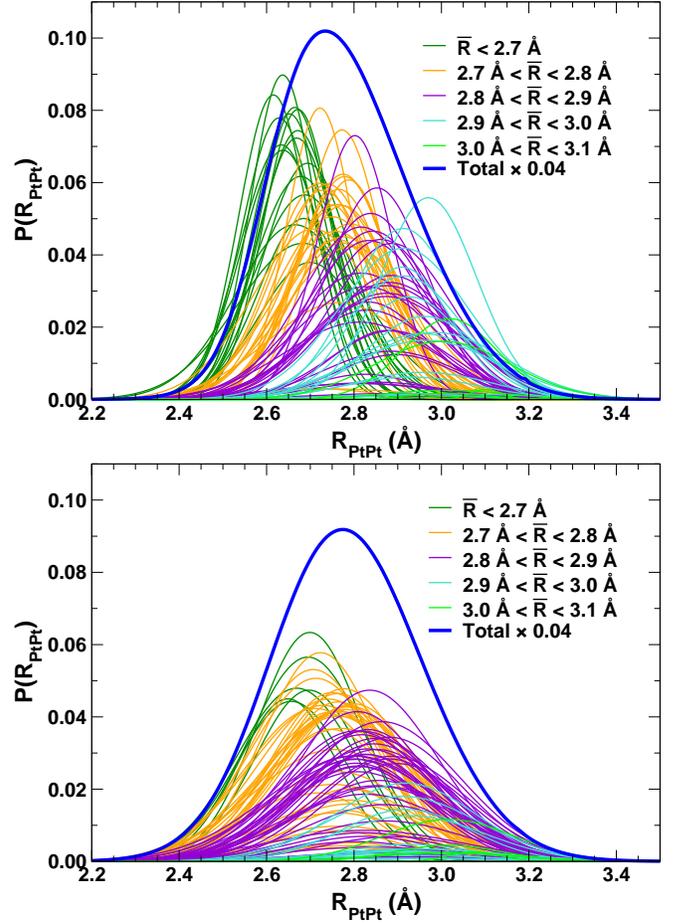

\includegraphics[scale=0.35,clip]{RDF_Decomp.Pt10Sn10.298.eps}
\includegraphics[scale=0.35,clip]{RDF_Decomp.Pt10Sn10.598.eps}
\caption{\label{fig:gri_r}
Decomposition of the total pair distribution function of Pt$_{15}$Sn$_{5}$ on
$\gamma$-Al$_{2}$O$_{3}$ at 298 K (top) and 598 K (bottom) into individual pair
components. The different colors label the mean distance for the pair and the
height of the distributions indicate their relative weight. Much of the width
and asymmetry of the near-neighbor distribution arises from cluster
inhomogeneity.
}
\end{figure}

\begin{figure}[ht]
\includegraphics[scale=0.35,clip]{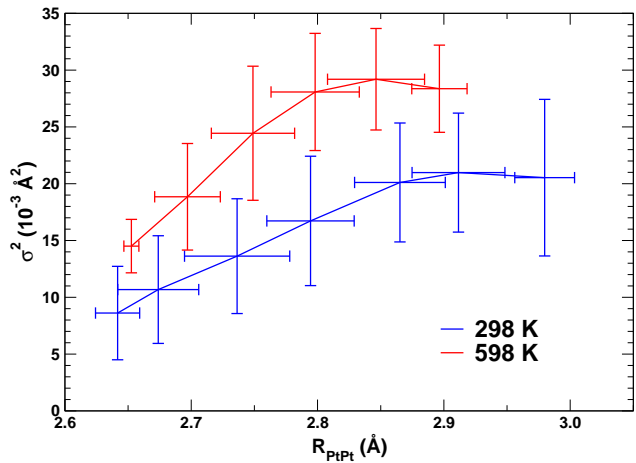}
\caption{\label{fig:s2vr}
Correlation between the mean Pt-Pt bond distances and their associated mean
MSRDs for Pt$_{10}$Sn$_{10}$ on $\gamma$-Al$_{2}$O$_{3}$ at 298 K and 598 K.
Both temperatures show similar behavior with weaker/longer bonds associated
with larger MSRDs.
}
\end{figure}

Note, however, that the quantities $\bar R$ and $\sigma^2$ refer to averages
over the entire cluster
and thus can give a misleading picture of the structure of inhomogeneous
systems. This difference is important in the Pt nano-clusters, for which
the mean Pt-Pt bond lengths and fluctuations depend on the local
environment and thus their locations inside the cluster.
Simulations of $\sigma^2$ can be carried out in terms of the distribution
of (partial) PDFs at each site $g(R)=\sum_i g_i(R)$.
The local near-neighbor distributions $\tilde g_i(R)$ have
net weights $\tilde g_i = \int dR\, \tilde g_i(R)$,
mean near neighbor bond lengths $\bar R_i = \int dR\, R\, \tilde g_i(R)$,  with
fluctuations  $\sigma_i^2 = \int dR\, (R-\bar R_i)^2 \tilde g_i(R)$.
Consequently the global average near-neighbor bond distance
$\bar R$ and MSRD $\sigma^2$ corresponding to experimental measurements
are
\begin{equation}
\bar R = \sum_i \bar R_i \tilde g_i, \  \qquad
\sigma^2 =  \bar\sigma_{D}^2 + \bar\sigma_{V}^2,
\end{equation}
where the local bond fluctuations are
$\bar\sigma_{V}^2 = \sigma_i^2 \tilde g_i$,
and the mean-squared disorder due to cluster inhomogeneity
is $\bar\sigma_D^2 = \sum_i (\bar R_i - \bar R)^2 g_i$.
Interestingly both of these contributions have DSD contributions from
low frequency fluctuations.
This procedure is illustrated in Fig.\ \ref{fig:gri_r} 
 which shows the decomposition of $\tilde g(R)$ vs mean bond length.
In particular, much of the width of the first neighbor peak in $\tilde g(R)$
comes from cluster inhomogeneity, and thus explains
the anomalously large disorder observed for the nanoclusters.\cite{kang2006}
Moreover, the individual PDFs in Fig.\ \ref{fig:gri_r} are not randomly
distributed, suggesting a
correlation between the mean Pt-Pt bond length and the associated MSRD of the
bond. This correlation is clearly shown in Fig.\ \ref{fig:s2vr}, with an increase
in the MSRD as the Pt-Pt bonds get longer/weaker.

\begin{figure}[ht]
\includegraphics[scale=0.35,clip]{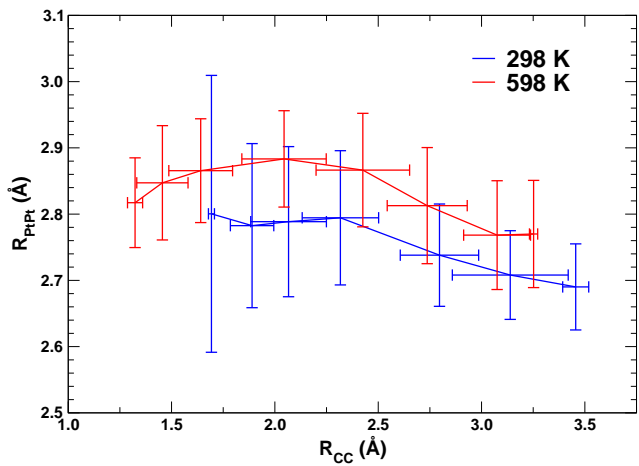}
\caption{\label{fig:rvcc}
Correlation between the Pt-Pt mean bond distances and their distance to the
center of
the nanoparticle (R$_{CC}$) for Pt$_{10}$Sn$_{10}$ on $\gamma$-Al$_{2}$O$_{3}$
at 298 K and 598 K. Both temperatures show similar behavior with shorter bonds
near the surface of the particle.
}
\end{figure}

Another example of structural correlations is shown in Fig.\ \ref{fig:rvcc},
which depicts the mean Pt-Pt bond distance $R_{PtPt}$ as a function of its
distance to the center of the nanoparticle ($R_{CC}$). Bonds near the surface of the
cluster ($R_{CC} \gtrsim$ 2.5\AA) are shorter, as expected from their reduced
number of near neighbors. This
threshold
is
in good agreement with
previous results\cite{vila2013} showing that the interior-surface
transition zone occurs between 2.5 and 3 \AA.


The low frequency DSD modes in Eq.\ (3) are particularly important
in these calculations, since they give large contributions to the MSRD
\begin{equation}
\sigma^2 = \frac{\hbar}{2\mu} \int d\omega\,
{\frac{\rho(\omega)}{\omega}}
\coth(\beta\hbar\omega/2)
\approx \frac{kT}{\mu} \int d\omega \frac{\rho(\omega)}{\omega^2},
\end{equation}
where the expression on the right is the high temperature limit. Here
$\mu$ refers to the reduced mass of a given bond pair, e.g, $M_{Pt}/2$
for Pt-Pt bonds. Thus it should be possible to distinguish DSD
from conventional vibrations by filtering the density of
modes $\rho(\omega)$, as shown in Fig.\ \ref{fig:vibdis}.
Similarly the contribution from each site $\sigma_i^2$ can be 
obtained by replacing $\rho(\omega)$ with the projected density of
vibrational modes $\rho_i(\omega)$. In this way, one can analyze
the net MSRD into contributions from different positions in the cluster
and their vibrational and DSD components.  A similar expression can be used
to calculate
the librational fluctuations of the CM, with $\mu$ replaced by $M_{CM}$
and $\rho$ the projected density of librational modes.
Note that this expression implies that $\sigma^2$ for the 
Pt-Pt bonds is linear in $T$ at high temperatures with a slope depending
on the inverse second moment of $\rho(\omega)$. 
Estimates of $\rho(\omega)$ can be obtained in various ways. For example
one approach is to use DFT/MD calculations
of $\langle \sigma^2 \rangle $, in real time over
a sufficiently long interval $\tau$,
\begin{equation}
\langle \sigma^2 \rangle = \frac{1}{\tau} \int_0^{\tau} dt\, \sigma^2(t) = 
 \int d\omega \, \sigma^2(\omega),
\end{equation}
where $\sigma^2(\omega) = \rho(\omega)/\mu\omega^2$.  Such real-time
calculations naturally account for the transient coupled-oscillator
motion and effects of charge fluctuations on the dynamics of the cluster,
as illustrated in Fig.\ \ref{fig:vibdis} (bottom).
Alternatively $\rho_R(\omega)$ can be obtained from
the FT of real-time DFT/MD trajectories or from
equation of motion techniques\cite{PhysRevB.85.024303}
\begin{equation}
 \rho_R(\omega) = \frac{2}{\pi} \int_0^{t_{max}} \langle Q_R(t)|Q_R(0)\rangle
\cos(\omega t) e^{-\epsilon t^2}.
\end{equation}

\begin{figure*}[th]
\includegraphics[scale=1.35,clip]{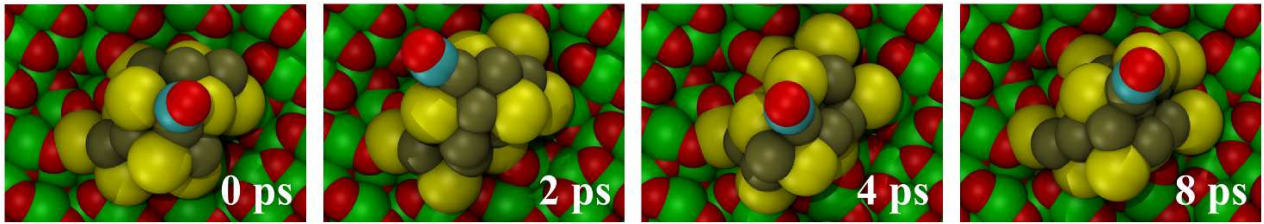}
\caption{\label{fig:Ads_CO_Snapshots}
Structural snapshots of the dynamics of CO on Pt$_{10}$Sn$_{10}$ on
$\gamma$-Al$_{2}$O$_{3}$ at 598 K. Here the sphere colors represent:
red, oxygen; blue, carbon; green, aluminium; yellow, tin; brown, platinum.
These snapshots show that the CO molecule
remains tethered to a Pt atom through its C-end for most of the simulation time.}
\end{figure*}

\section{Nanoscale Reaction Rates}
\label{sec:nanorates}

Formally, calculations of catalytic properties such as reaction rates
depend on free energies from equilibrated statistical ensembles
that contain all accessible regions of phase space. The effects of disorder\cite{zwanzig1990rate,zwanzig:3587} and
the presence of a variety of possible transition
paths\cite{bolhuis2002transition,dellago1998efficient} have been addressed in the past. Due to its simplicity, however, 
transition state theory (TST) is still the most widely used approach
for rate calculations in molecular and nanoscale systems.\cite{truhlar1996} This approach is usually based on time-independent 
nanoparticle potential energy surfaces and the
assumption of quasi-equilibrium between reactants and transition state.
In contrast, 
our SRR model based on finite-temperature DFT/MD calculations
avoids many of these difficulties.  We stress that {\it ab initio} DFT
is crucial for many properties, as classical MD simulations with
model potentials do not adequately
capture charge fluctuations, bond-breaking, diffusion and other
non-equilibrium effects.\cite{jonsson2000}  Moreover DFT/MD
is a good approach for non-equilibrium structural
and physical properties at high temperatures, since it builds
in anharmonic and structural disorder.

 Statistical thermodynamic arguments similar to those in Sec.\ II.\
can be applied to nano-scale reaction rates. Our analysis
suggests that it may be
important to treat the behavior of reactant molecules on the dynamically
fluctuating surface of a nano-cluster, e.g., with our SRR model,
rather than on selected static structures.
 Thus, the
slow fluctuating
modes of both the cluster and the reactant molecules
can couple,
resulting in an statistical ensemble of reaction barriers and
reaction rates.

The SRR catalysis model thus consists of a system of reactant
molecules in contact with a fluctuating cluster that can exchange
energy to and from it. Our treatment below focuses on
the rate limiting step of crossing the transition barrier; more detailed
approximations should take into account the diffusion rates of the
molecules on the surface of the nano-clusters to the reaction site.
Given that diffusion processes are dominated by low frequency
modes, such rates are likely influenced by the slow nature of the
DSD.  Using arguments analogous to
those of Sec. II, the probability distribution of reactant energies
$\epsilon$  is given by a statistical ensemble with a fluctuating
cluster energy distribution $E$ relative to the mean $\bar E$:
 $\Omega(E-\epsilon) \propto \exp[S(E) - \tilde \beta \epsilon] $
where $\tilde \beta = \partial S(E)/\partial E = 1/k\tilde T$ is the
fluctuating inverse local temperature at energy $E$.
Thus we obtain
\begin{equation}
 P(\epsilon) \approx \langle \Omega_R(\epsilon)
e^{- \tilde \beta \epsilon}
\rangle = \langle e^{-\tilde \beta f(\epsilon;\tilde T)} \rangle,
\end{equation}
where $\Omega_R(\epsilon)$ is the configurational entropy of the reactant
molecules on the cluster, and the average is carried out over the
distribution of cluster energies $E$ and hence values of $\tilde\beta$.
This result shows why the average over terms
with $\tilde\beta$ is not the same as that with fixed $\beta$ when
temperature fluctuations are large.

Consequently, calculations of reaction rates $\kappa$  for the transition
$[A]+[B] \rightarrow [AB^*] \rightarrow [P] $ require a generalization
of transition state theory to the case with fluctuating barriers,
\begin{equation}
\kappa =  \frac{kT}{h} \left\langle
\frac{\tilde Z^*_{AB}}{\tilde Z_A \tilde Z_B}
e^{- \tilde \beta \epsilon_b} \right\rangle
\end{equation}
Here $h$ is  Planck's constant and the averages in the partition functions
are carried out over all nano-cluster configurations:
$Z^*_{AB}$ is the partition function of
the activated complex, and $Z_A$ and $Z_B$ the partition functions of
the reactants $A$ and $B$, and again the average is carried out over
the range of local temperatures $\tilde\beta$.
This result is analogous to the usual Arrhenius law, but with a dynamically
averaged ``attempt frequency'' prefactor and
fluctuating reaction barriers $\langle e^{-\tilde\beta\epsilon_b}\rangle$.
Both of these factors are affected by DSD.
It is easy to see, for example, using a first order
cumulant expansion, why the latter average is always greater than
the reactivity at the mean barrier $\langle\epsilon_b\rangle$
\begin{equation}
\langle e^{\tilde\beta\epsilon_b}\rangle =
e^{-\tilde\beta\langle \epsilon_b\rangle}
e^{(+1/2) \tilde\beta^2 \sigma^2_{\epsilon_b}} 
\geq e^{-\tilde\beta\langle \epsilon_b\rangle}.
\end{equation}
The average leads to a reduced temperature dependent ``effective barrier"
$ \epsilon'_b = \epsilon_b - (1/2) \tilde \beta \sigma^2_{\epsilon_b},$
which is comparable to the lowest barriers
typically encountered during reactions. Note, however, that
the cumulant estimate
assumes small fluctions in $\epsilon_b$, more generally
the inequality is always valid and the lowest barriers dominate.

Approximate calculations of the molecular partition functions 
in these fluctuating TST models can be carried out in
terms of the {\it local free energies} of the molecules bonded to
the fluctuating  cluster or at the transtion state
using a generalization of the
coupled-oscillator model discussed above.  One expects that
the low frequency
modes
of the reacting molecule
will be strongly coupled to the cluster and depend both
on the instantaneous orientation and local geometry of the molecule
on the cluster surface at a given time. This coupling is expected
to be transient and it is plausible that cluster fluctuations will
assist a molecule in probing a wider range of degrees of freedom
than on a static support.
Within this model, the local free energy of an adsorbed molecule with
position $\vec r$ is given by
\begin{equation}
 f(\tilde T) = \epsilon_0(\vec r) + 
 k \tilde T \int d\omega\, \rho_R(\omega)
\ln [2\sinh(\tilde \beta \hbar\omega/2)],
\end{equation}
where $\rho_R(\omega)$ is the projected local density of vibrational states
at a given reactant molecule at position $\vec r$ on
the surface of the cluster and
$\tilde T$ is the effective temperature of the cluster.
As with the total free energy of the cluster, the local free energy
of the molecule $f(\tilde T)$ contains terms coupling to the fast
vibrational density of states of the cluster,
and to the relatively slow librational motion of the CM and other
contributions to DSD.  Since reactant molecular vibration 
frequencies are often large compared to thermal motion
($\hbar\omega \gg kT$), their vibrational states
are typically frozen in their ground state, and the high temperature limit is
inappropriate.  On the other hand, relatively low energy bending and
rotational modes may be thermally active and contribute strongly.
As with DSD, one expects that the MSRD $\sigma^2$ of the
reactant molecule will contain contributions from the low
frequency librational modes, even for weak couplings as contributions
to $\sigma^2$ are proportional to $1/\omega^2$
(cf.\ Fig.\ 5 of Ref.\ [\onlinecite{PhysRevB.76.014301}])
Thus it is possible that these low frequency fluctuations can
significantly increase the entropy fluctuations and hence the
catalytic activity of reactants, though quantitative calculations
will be needed to assess the net effect.

\section{Model
Calculations}

\subsection{DSD and adsorbate dynamics}

As an illustration of the theory discussed above, we present calculations
and sample reaction paths for prototypical molecules adsorbed on
nano-clusters.
In order to address the effects of DSD on their catalytic activity 
we first address the ``simpler'' problem of adsorbate dynamics on
the surface of the particles.
For this purpose we have carried out MD simulations
similar to those described in the previous sections, but with the addition of an
adsorbate molecule (CO) on the surface of the nanoparticle. Fig.
\ref{fig:Ads_CO_Snapshots} shows a series of snapshots illustrating
the high mobility of the molecule over the surface.

Qualitatively, while the CO molecule position is strongly coupled to the DSD of
the nanoparticle, its internal structure is not expected to depend
strongly on disorder. The MD simulations support this statement: First, the
CO molecule is tethered to a Pt atom through its carbon-end for most of
the simulation, so its position is modulated by the slow Pt motion.
Nevertheless, brief
incursions into bifurcated anchoring states are also observed. These noticeably
affect the internal structure of the adsorbate by elongating the
CO bond. Second, the low frequency dynamics associated with the DSD are likely
to couple better with the low frequency C-Pt bonds, rather than with the high
frequency adsorbate covalent bonds.

\begin{figure}[t]
\includegraphics[scale=0.35,clip]{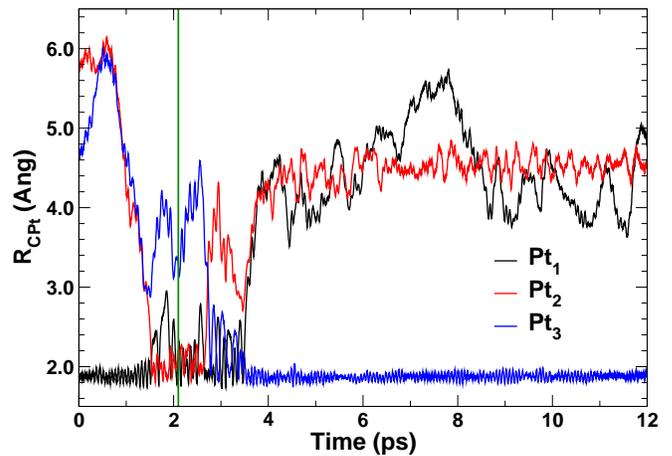}
\caption{\label{fig:rCPt_Traj1}
Dynamics of the three shortest R$_{CPt}$ bonds in a typical trajectory for CO
adsorbed on Pt$_{10}$Sn$_{10}$ on $\gamma$-Al$_{2}$O$_{3}$ at 598 K. The
vertical green line indicates the end of the thermalization stage. The region
between $\sim$2-4 ps shows a quick transitions between Pt anchoring points,
with brief incursions into bifurcated anchoring states. These states have a
clear effect on the R$_{CO}$ distance (see Fig.\ \ref{fig:Decomp_rCOm}).}
\end{figure}

\begin{figure}[ht]
\includegraphics[scale=0.35,clip]{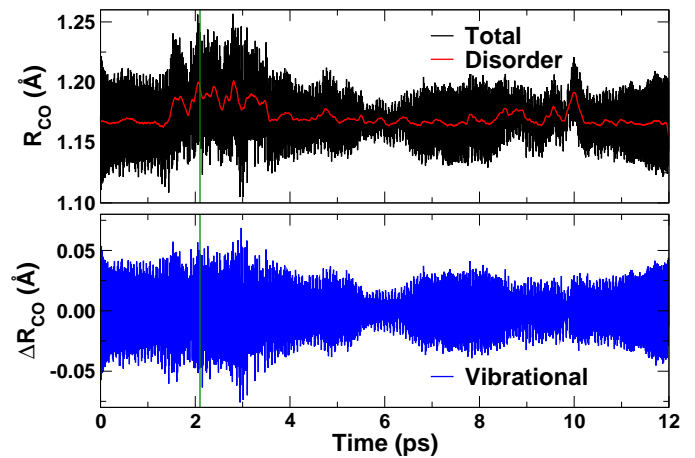}
\caption{\label{fig:Decomp_rCOm}
Decomposition of the R$_{CO}$ dynamics into vibrational and disorder components
for CO adsorbed on Pt$_{10}$Sn$_{10}$ on $\gamma$-Al$_{2}$O$_{3}$ at 598 K. The
vertical green line indicates the end of the thermalization stage.}
\end{figure}

To investigate this coupling between the adsorbate dynamics and the
nanoparticle DSD more quantitatively, we have have filtered
out the slow, stochastic component from the intra-adsorbate (R$_{CO}$) and
adsorbate-particle (R$_{CPt}$) dynamics, similar to that
in Fig.\ \ref{fig:vibdis}. The fast
transition between Pt anchoring points described above can be clearly seen in a
typical trajectory (Fig.\ \ref{fig:rCPt_Traj1}).
Here the region between $\sim$2-4 ps shows the brief incursions into bifurcated
anchoring states. These transient states have a clear effect on the internal
structure of the adsorbate, as seen in Fig.\ \ref{fig:Decomp_rCOm},
which shows the decomposition of the R$_{CO}$ dynamics into vibrational and disorder
components. The transient mean bond distance (represented by the red
``Disorder'' curve) is increased in the bifurcated bond, while the amplitude of
the vibrational component (blue ``Vibrational'' curve) is enhanced due to the
weakened bond.

\begin{figure}[ht]
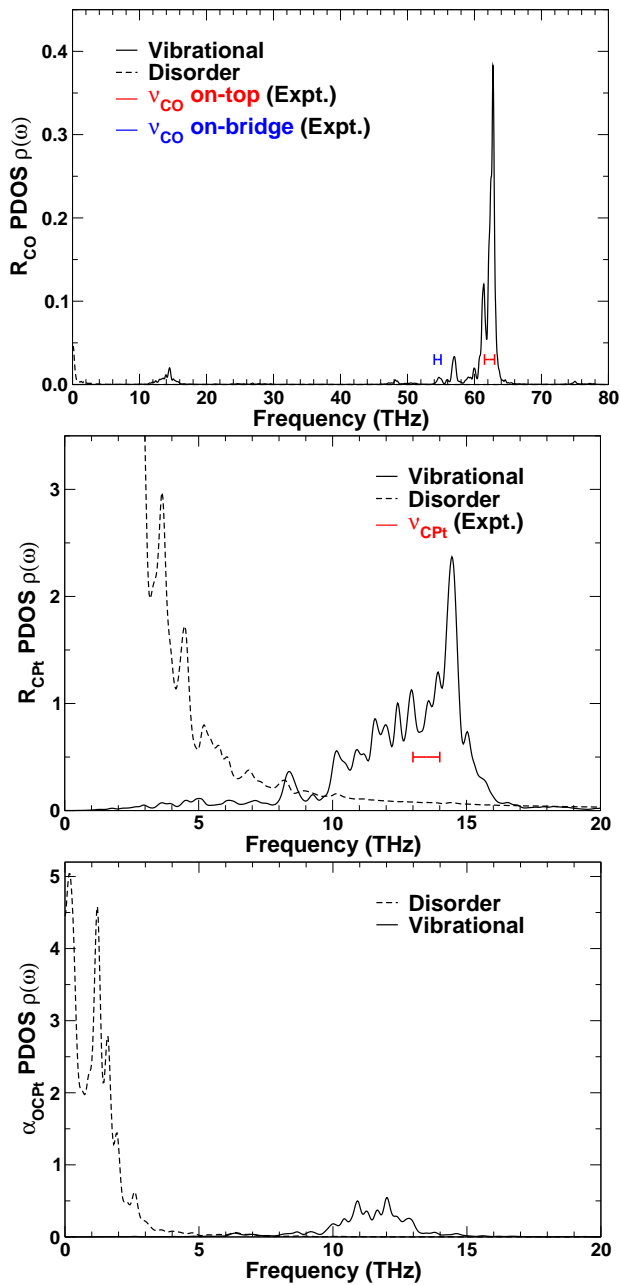

\includegraphics[scale=0.33,clip]{vdos_rCOm.eps}
\includegraphics[scale=0.33,clip]{vdos_rCPt.eps}
\includegraphics[scale=0.33,clip]{vdos_aOmCPt.eps}
\caption{\label{fig:vdos}
Average projected density of states for the vibrational and dynamic structural disorder
components of the R$_{CO}$ (top), R$_{CPt}$ (middle) and
$\widehat{PtCO}$ ($\alpha_{OCPt}$, bottom) trajectories in
Pt$_{10}$Sn$_{10}$ on $\gamma$-Al$_{2}$O$_{3}$ at 598 K.
The experimental CO stretch frequency
for on-top adsorption sites ranges from 61.5 to 63 THz, while for on-bridge
sites it ranges from 54 to 55 THz. For the C-Pt stretch, the frequency
ranges from 13 to 14 THz.
}
\end{figure}

The projected density of states $\rho(\omega)$ for each component of these
trajectories can be calculated from the autocorrelation function
$\langle R(0)\cdot R(t)\rangle$ of the coordinate of interest using
\begin{equation}
 \rho(\omega) = \int_0^{\infty} \langle R(0)\cdot R(t)\rangle
\cos(\omega t) e^{-\epsilon t^2}.
\end{equation}
These spectra are shown in Fig.\ \ref{fig:vdos} (top), where
$\epsilon$ is a broadening parameter added for convenience. The
most prominent features correspond to the CO stretch for both on-top (between 60
and 64 THz) and on-bridge (small peaks between 54 and 58 THz). These frequency
ranges
are in good agreement with the experimental
values.\cite{Kung2000L627,McCrea2001238,mccrea2002,hu2012} Also noticeable is
a small
feature between 12 and 16 THz associated with the molecule-particle
interaction, as discussed below in further detail. As
expected, the negligible weight of the disorder confirms that
the intramolecular dynamics of CO are largely independent of DSD.

The disorder has a much larger role in the molecule-nanoparticle 
dynamics, as seen in Fig.
\ref{fig:vdos} (middle), where the R$_{CPt}$ modes below 5 THz are highly
coupled to the
DSD. The vibrational component shows a prominent
feature between 13 and 15 THz, in reasonable agreement with the experimental
range of 13-14 THz for the C-Pt stretch mode.\cite{paffet1990} In addition,
the R$_{CPt}$
dynamics show a broad feature between 10 and 13 THz which results from strong
coupling to the $\widehat{PtCO}$ bending, as seen in
Fig.\ \ref{fig:vdos} (bottom).
These results point to the importance of nanoparticle fluctuations on
the pre-dissociation regime where the reacting molecule both explores a
variety of surface sites, and the physical properties of the sites themselves
vary dynamically.


\subsection{DSD in reaction barriers}

\begin{figure}[ht]
\includegraphics[scale=0.35,clip]{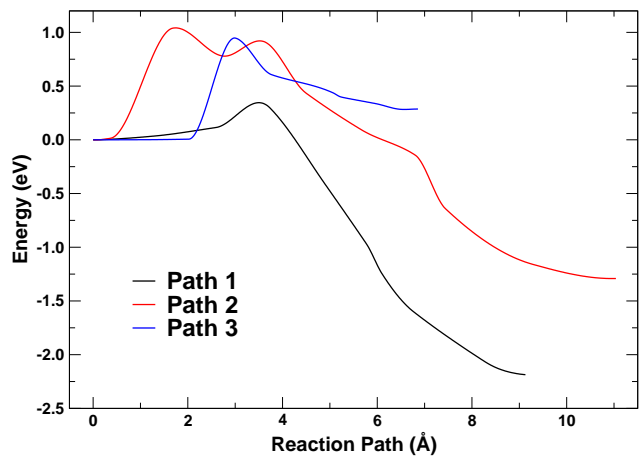}
\caption{\label{fig:PES_All}
Typical NEB reaction paths for the dissociation of O$_2$ on fully relaxed
Pt$_{10}$Sn$_{10}$ supported on $\gamma$-Al$_{2}$O$_{3}$. Paths energies
have been shifted to match at the reactant state. The reaction path coordinate
corresponds to the integrated displacement of all atoms in the system.
}
\end{figure}

As discussed in Sec.\ \ref{sec:nanorates}, the dynamic disorder creates an
effective activation barrier $\epsilon'_b \leq \epsilon_b$
due to barrier fluctuations. To study these fluctuations, we performed
nudged elastic band (NEB) calculations of the dissociation of O$_2$ on
Pt$_{10}$Sn$_{10}$ supported on $\gamma$-Al$_{2}$O$_{3}$. Fig.
\ref{fig:PES_All} shows results for three sample reaction paths, generated for
nanoparticle conformations with different DSD, sampled from DFT/MD trajectories.
These paths represent rather different behaviors, including exothermic and
endothermic thermodynamics, and the presence of a high energy intermediate
in path 2.
From these results we can roughly estimate the mean barrier ($\epsilon_b \simeq
$ 0.8 eV) and barrier fluctuations ($\sigma_{\epsilon_b} \simeq$ 0.5 eV). In
addition, we can estimate the nanoparticle distortions induced during the
dissociation process by analyzing the atomic displacements along the reaction
path.
For the paths shown in Fig.\ \ref{fig:PES_All}, the metal atoms
directly bound to the O$_2$ molecule experience displacements of the
order of 0.2 \AA\ along the reaction path.
Such nanoparticle distortions are of similar magnitude to those observed due
to DSD (0.2-0.4 \AA). The remainder of the nanoparticle also participates
in the reaction, but with smaller displacements of O(0.1 \AA).

\begin{figure}[ht]
\includegraphics[scale=0.35,clip]{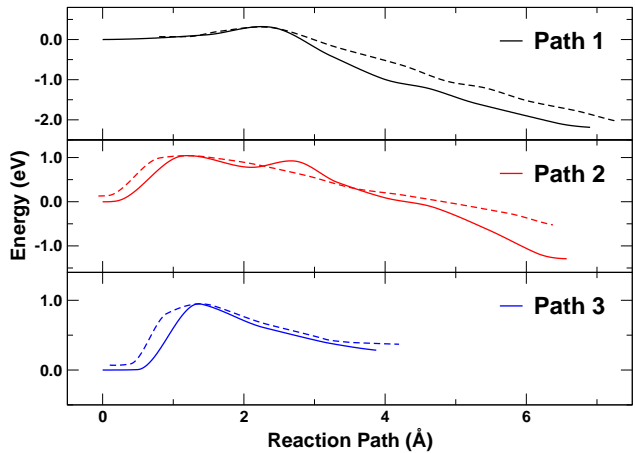}
\caption{\label{fig:PES_All_Smp}
Comparison of the fully relaxed NEB reaction paths for the dissociation of
O$_2$ on
Pt$_{10}$Sn$_{10}$ supported on $\gamma$-Al$_{2}$O$_{3}$ (full lines) with the
reaction paths obtained by fixing the surface and nanoparticle structure at the
transition state conformation. The energies for the fully relaxed paths
have been shifted to match at the reactant state. The reaction path coordinate
corresponds to the integrated displacement of all relaxed atoms in the system.
}
\end{figure}

The effects of DSD on the reaction barriers are twofold: First, direct effects
change the chemical nature of the reactants, transition states and products.
These direct effects are usually visible as large energy changes, such as the
shift from endothermic to exothermic behavior in paths 3 and 2, respectively,
and are
driven by major rearrangements of the structure of the nanoparticle and its
interaction with the adsorbate. Second, smaller fluctuations in structure that
do not change the chemical nature of the different states yet still induce
changes
in the activation barrier. To quantify these indirect effects we have also
computed reaction barriers where all atoms, except those in the dissociating
O$_2$ molecule, are fixed at the transition state structure of the paths in
Fig.\ \ref{fig:PES_All}. These rigid nanoparticle paths are shown in
Fig.\ \ref{fig:PES_All_Smp}.
Overall, these paths are smoother than the fully relaxed ones.
For instance, the nanoparticle fluctuations induce a variety of minima and
shoulders in the reaction path. As expected, the reaction barriers are higher
for the fully relaxed paths. The difference is only 0.1-0.2 eV; but the effect
on the exponential character of reaction rates is nevertheless substantial.
This result demonstrates our premise that DSD in nano-structures
can significantly affect reaction rates.


\section{Conclusions}

We have investigated the effects of dynamic structural disorder on the
properties of supported metal nanoparticles, from both their morphological and
reactivity perspectives. Our results suggest that a real-time approach
that accounts for their fluctuating bonding and electronic structure
may provide a useful approach to better understand of their catalytic activity. In particular, we
have found that the dynamic structural disorder (DSD) in nanoclusters leads to a
larger statistical ensemble of configurations than on solid surfaces.
The \textit{shake, rattle and roll} (SRR) concept makes it possible to
simulate 
the statistical ensemble of possible reaction sites efficiently, compared to
the effort needed for enumerating structures and Boltzmann factors.
This can reveal both the surface structure and
dynamics, which can be more important than the global average morphology.
Since experimental probes measure global averages, it is important in the
analysis to differentiate between surface and internal structure. The local
electronic structure of the binding sites is highly heterogeneous. For instance,
the oxidation state of a Pt atom varies significantly depending on its local
environment. Thus, the DSD in these systems effectively induces additional
active sites. Our results highlight the importance of large 
fluctuations driven by the stochastic motion of the center of mass of the
nanoparticles and their transient bonding to the support. These fluctuations
affect the internal energy distribution as well as the structure of the
nanoparticles, thus having a measurable effect on both the properties observed
by experimental probes like EXAFS, as well as the observed reaction rates.

\begin{acknowledgments}
The authors wish to thank S. R. Bare, J. J. Kas, S.D. Kelly, M. Tromp,
and especially A. Frenkel for advice and useful references.
This work was supported by DOE  Grant DE-FG02-03ER15476
with substantial computer support from DOE-NERSC.
 \end{acknowledgments}

\bibliographystyle{apsrev}

\end{document}